\documentclass[prl,twocolumn,superscriptaddress,groupedaddress]{revtex4}  

\usepackage{graphics}  
\usepackage{dcolumn}   
\usepackage{bm}        
\usepackage{amssymb}   
\usepackage{epstopdf}
\usepackage{epsfig}
\usepackage{subfig}
\usepackage{float}
\usepackage{array}
\usepackage{amsmath}
\usepackage{braket}
\usepackage{verbatim}

\usepackage{titlesec} 
\titleformat{\section}[block]{\bfseries}{Sec.\arabic{section}}{1em}{ }[]
\titleformat{\subsection}[block]{\itshape\mdseries}{\arabic{section}.\arabic{subsection}}{1em}{}[]

\usepackage[english,francais]{babel}

\begin{document}

\bibliographystyle{unsrt}

\date{29 January 2019}

\title{Spontaneous Long-range Vortex-antivortex Pair in the Two-band 2D Superconductor}

\author{Zhi-Wei Wang \\ email: bishop@i.shu.edu.cn } \affiliation{ Department of Physics, Shanghai University, 200444 Shanghai, People’s Republic of China }

\begin{abstract}
The relaxation iterative method is used to minimize Ginzburg-Landau model for the two-band superconductor with Josephson-coupling. A stable spontaneous vortex-antivortex pair with long range order has been revealed. Our result appears due to the Josephson-coupling effect, which leads to a linearly-dependence of total free energy on the distance between vortex-antivortex pair, hence accounts for this phenomenon.\\

\end{abstract}

\maketitle

\section{\label{sec:level1}1.Introduction}
Traditionally, according to the Mermin-Wagner (MW) theorem, it's commonly believed that no continuous spontaneous symmetry breaking and therefore no phase transition (except the discrete situation) could occur  in an isotropic two-dimensional system without topological defect. Then, Korsterlitz and Thouless\cite{kosterlitz1973ordering}, also by Berezinskii\cite{berezinskii1971destruction}, has indicated the existence of topological transition in  a 2D system which explains the existence of vortex-antivortex pair observed in the superfluid helium-3. However, this type of phase transition cannot occur in a charged system (superconductor) since the repulsive energy between two vortices with opposite circulation fall off like $r^{-1}$ instead of the necessarily ln r. In order to get a stable vortex-antivortex pair in superconductor, quantum confinement effect has been introduced  where the vortex-antivortex pair can be stabilized with the reflection of boundary (namely: the mirror reflection force )\cite{misko2003stable}. Also, the possibility to stabilize the vortex-antivortex pair in Superconductor with the non-zero demagnetization factor has been discussed by Epstein et al.\cite{beasley1979possibility}. Moreover,  vortex-antivortex pair can appear by quantum magnetic dots pinning discussed by $Milo\breve{s}evi\acute{c}$\cite{milovsevic2004vortex}. Recently, Timm and Appel \cite{timm1994vortex}  has indicated that the total free energy of system in the two-layered superconductor with Josephson coupling is linearly-dependent on their separation and argued that KT's theorem, which assumes a logarithmic interaction, is not applicable to such a two-layered superconductor with Josephson-coupling, therefore, suggests the possibility of a spontaneous long-range vortex-antivortex excitation in the charged system.   \\

\vspace{-0.5cm}

Multi-band superconductivity\cite{zehetmayer2013review} in different systems such as $ MgB_2 $  \cite{iavarone2002two}  \cite{buzea2001review}and iron-based materials  has been confirmed in the last two decades by several experimental techniques, showing quite different thermodynamic, reversible mixed-state and other abnormal properties from one-band superconductor due to interband coupling. To our knowledge, although numerous research has been done on two-band model, like  phase soliton Lin et al.\cite{lin2012phase}, (while the stability of their results needs further discussion)and by Tanaka\cite{tanaka2001soliton}, Type-1.5 Superconductivity  by Moshchalkov et al.\cite{moshchalkov2009type}, a spontaneous vortex-antivortex pattern has not  been reported. In present work, a novel long-range vortex-antivortex pair has been found by minimizing the free energy based on two-band Ginzburg-Landau model with Josephson-coupling where the direct iterative relaxation  method  \cite{wiki:xxx} \cite{wang1991numerical} \cite{adler1984relaxation}has been used with a periodic boundary condition.  Due to the Josephson term, the interband symmetry has been broken, so wavefunction will organize itself under suitable initial condition.

\section{2.Model}

 We start with the two-band Ginzburg-Landau model with Josphson-coupling which provides the parameter accessible to experiments. Derived from  microscopic BCS theory with Gor'kov's Green's function technique, the GL free energy in such a model can be written as  \cite{zehetmayer2013review}  \cite{vagov2012two}

\begin{widetext}

\begin{eqnarray}
      F=\frac{1}{S}\int dx \int dy \sum\limits_{j=1,2} \left[\alpha _j|\psi _j|{}^2+\frac{\beta _j}{2}|\psi _j|{}^4+\frac{1}{2m_j}\left|(-i \hbar \nabla  -\frac{e^*}{c} A)\psi _j\right|{}^2\right]+\frac{1}{8\pi }(\nabla \times \mathbf{A})^2+\gamma (\psi _1^*\psi _2+\psi _2^*\psi _1)
\end{eqnarray}

Here,wavefunction $\psi_j$ was introduced as a complex order parameter.$|\psi_j|^2$ was to represent the local density of superconducting electrons $n_s(r)$. $m_j$ is electron effective mass, pA is the magnetic vector potential, $\gamma$ is the strength of Josephson-coupling and j indicates the band index from 1 to 2.The variational of F with respect to the $\psi_j $ and A leads to the usual GL equation.\\

\ To apply the iteration method, two steps are necessary. The first step is to make the free energy dimensionless by scaling energies by the superconducting condensation energy of band one, $\alpha_1^2/\beta_1$ and lengths by the coherence length of band one $\xi_1 $=$ (-\hbar^2 / 2 m_1 \alpha_1)^{1/2} $ ,
setting also $\psi_1=\widetilde{\psi_1} \psi_{1 \infty} $, $\psi_2=\widetilde{\psi_2} \psi_{2 \infty} $ where $ \psi_{1 \infty}^2=-\alpha_1/\beta_1, \psi_{2 \infty}^2=-\alpha_2/\beta_2 $ and $A=\frac{\Phi_0}{2\pi \xi_1} \mathcal{A}$

Equation becomes

\begin{equation}
\begin{split}
   \mathcal{F} &=\frac{1}{S}\int dx \int dy  -|\widetilde{\psi _1}|^2+\frac{1}{2}|\widetilde{\psi _1}|^4+ |(\nabla- i \mathcal{A})\widetilde{\psi _1} |^2 -a_2          |\widetilde{\psi _2} |^2+\frac{b_2}{2} |\widetilde{\psi _2}|^4+c_2|(\nabla- i \mathcal{A})\widetilde{\psi _2} |^2+\kappa^2 (\nabla \times \mathcal{A})^2  \\
   &+d_2 (\widetilde{\psi _1^*} \widetilde{\psi _2}+\widetilde{\psi _2^*}\widetilde{\psi _1})
\end{split}
\end{equation}

where $a_2=b_2=\frac{\alpha_2^2}{\alpha_1^2} \frac{\beta_1}{\beta_2}$,
$c_2=\frac{m_1}{m_2}\frac{\alpha_2}{\alpha_1} \frac{\beta_1}{\beta_2}$,
$d_2=\frac{\gamma}{|\alpha_1|} (\frac{\alpha_2}{\alpha_1} \frac{ \beta_1}{\beta_2})^{0.5}$,
$\kappa=\frac{m c}{e^* \hbar} (\frac{\beta}{2 \pi })^{0.5} $.

The second step is to discretize  the GL free energy functional after variational, and therefore, by using direct local-optimization technique for a multivariable function, one can set the relaxation iteration formula as follow.

\begin{equation}\label{eqs1n1}
\begin{array}{l}
\widetilde{\psi_1}^{(n+1)}(i,j)=\widetilde{\psi_1}^{(n)}(i,j)-\epsilon \frac{\partial{   \mathcal{F}}}{\partial{\widetilde{\psi_1}^*(i,j)} }|^{(n)}  ;\qquad
\widetilde{\psi_2}^{(n+1)}(i,j)=\widetilde{\psi_2}^{(n)}(i,j)-\epsilon \frac{\partial{   \mathcal{F}}}{\partial{\widetilde{\psi_2}^*(i,j)} }|^{(n)}\\

\mathcal{A}_x^{(n+1)}(i,j)=\mathcal{A}_x^{(n)}(i,j)-\epsilon \frac{\partial{   \mathcal{F}}}{\partial{\mathcal{A}_x(i,j)} }|^{(n)}           ;\qquad
\mathcal{A}_y^{(n+1)}(i,j)=\mathcal{A}_y^{(n)}(i,j)-\epsilon \frac{\partial{   \mathcal{F}}}{\partial{\mathcal{A}_y(i,j)} }|^{(n)}\\

\end{array}
\end{equation}

Here, the original area  S= $ L_x \times L_y $ is re-scaled into lattice area $N_x \times N_y $($ L_x=N_x d_x,L_y=N_y d_y $). And n stands for the generation of iteration steps, with a suitable choice of factor $\epsilon $ and suitable initial state,the $\mathcal{F}$ will reach its optimal state monotonically.

\begin{equation}\label{eqs1n1}
\begin{array}{l}

 \frac{\partial{   \mathcal{F}}}{\partial{\widetilde{\psi_1}^*(i,j)} }=-\widetilde{\psi_1}(i,j)+|\widetilde{\psi_1}(i,j)|^2\widetilde{\psi_1}(i,j)-(\nabla -i\mathcal{A}(i,j) )^2 \widetilde{\psi_1}(i,j)+d_2 \widetilde{\psi_2}(i,j)  \\

\frac{\partial{   \mathcal{F}}}{\partial{\widetilde{\psi_2}^*(i,j)} }=-a_2 \widetilde{\psi_2}(i,j)+b_2 |\widetilde{\psi_2}(i,j)|^2\widetilde{\psi_2}(i,j)-c_2 (\nabla -i\mathcal{A}(i,j) )^2 \widetilde{\psi_2}(i,j)+d_2 \widetilde{\psi_1}(i,j)  \\

 \frac{\partial{   \mathcal{F}}}{\partial{\mathcal{A}_x(i,j) } }=
 i [\widetilde{\psi_1}^*(i,j) \nabla_x \widetilde{\psi_1}(i,j)-\widetilde{\psi_1}(i,j)\nabla_x \widetilde{\psi_1}^*(i,j) ]
 +c_2\ i [\widetilde{\psi_2}^*(i,j) \nabla_x \widetilde{\psi_2}(i,j)-\widetilde{\psi_2}(i,j)\nabla_x \widetilde{\psi_2}^*(i,j) ]          \\

 +2 \mathcal{A}_x(i,j)|\widetilde{\psi_1}|^2  +2 c_2 \mathcal{A}_x(i,j)|\widetilde{\psi_2}|^2
 +2\kappa^2 (\nabla \times \nabla \times \ \mathcal{A}(i,j))|_{along\ the\ x\ direction}   \\

 \frac{\partial{   \mathcal{F}}}{\partial{\mathcal{A}_y(i,j) } }=
 i [\widetilde{\psi_1}^*(i,j) \nabla_y \widetilde{\psi_1}(i,j)-\widetilde{\psi_1}(i,j)\nabla_y \widetilde{\psi_1}^*(i,j) ]
 +c_2\ i [\widetilde{\psi_2}^*(i,j) \nabla_y \widetilde{\psi_2}(i,j)-\widetilde{\psi_2}(i,j)\nabla_y \widetilde{\psi_2}^*(i,j) ]          \\
 +2 \mathcal{A}_y(i,j)|\widetilde{\psi_1}|^2  +2 c_2 \mathcal{A}_y(i,j)|\widetilde{\psi_2}|^2
 +2\kappa^2 (\nabla \times \nabla \times \ \mathcal{A}(i,j))|_{along\ the\ y\ direction}\\

\end{array}
\end{equation}

\end{widetext}

\vspace{-1cm}
where
\vspace{-1cm}

\begin{equation}\label{eqs1n1}
\begin{array}{l}
\frac{\partial{\widetilde{\psi}(i,j)}}{\partial{x}}=\frac{\widetilde{\psi}(i+1,j)+\widetilde{\psi}(i-1,j)}{2d_x}       \\
\frac{\partial^2{\widetilde{\psi}^2(i,j)}}{\partial{x^2}}=\frac{\widetilde{\psi}(i+1,j)-2\widetilde{\psi}(i,j)+\widetilde{\psi}(i-1,j)}{d_x^2}       \\
\frac{\partial{\widetilde{\psi}(i,j)}}{\partial{y}}=\frac{\widetilde{\psi}(i,j+1)+\widetilde{\psi}(i,j-1)}{2d_y}       \\
\frac{\partial^2{\widetilde{\psi}^2(i,j)}}{\partial{y^2}}=\frac{\widetilde{\psi}(i,j+1)-2\widetilde{\psi}(i,j)+\widetilde{\psi}(i,j-1)}{d_y^2}       \\
\end{array}
\end{equation}
It's easy to find out the boundary condition under the principle of gauge invariance \cite{doria1989virial}

   \begin{eqnarray}
&   \mathcal{A}(x+b_v)= \mathcal{A}(x)+ \nabla \chi_v(x)      \\
&\triangle(x+b_v)=\triangle (x) exp(i \frac{e^*}{\hbar c} \chi_v(x))
   \end{eqnarray}

   while the continuity of function must be satisfied as
   \begin{eqnarray}
\frac{e^*}{\hbar c} [\chi_\alpha(x)-\chi_\alpha(x+b_\alpha)+\chi_\beta(x+b_\alpha)-\chi_\beta(x)]=2\pi N \quad
   \end{eqnarray}
   Here N stands for an integer, Therefore the boundary condition can be written as follow. Here i/j stands for the lattice number of x direction and y direction from 1 to $N_x/N_y$, k stands for the imaginary unit and $\Phi$ stands for total reduced flux.

\vspace{-0.5cm}

\begin{equation}\label{eqs1n1}
\begin{array}{l}
\widetilde{\psi_1}(1,j)=\widetilde{\psi_1}(N_x,j)exp(-k\ j\ \Phi /N_x )         \\
\widetilde{\psi_2}(1,j)=\widetilde{\psi_2}(N_x,j)exp(-k\ j\ \Phi /N_x )           \\

\mathcal{A}_x(1,j)=\mathcal{A}_x(N_x,j)                \\
\mathcal{A}_y(1,j)=\mathcal{A}_y(N_x,j)-\Phi/(d_x\ N_x) \\

\widetilde{\psi_1}(i,1)=\widetilde{\psi_1}(i,N_y)exp(-k\ \Phi /2 )            \\
\widetilde{\psi_2}(i,1)=\widetilde{\psi_2}(i,N_y)exp(-k\ \Phi /2 )            \\

\mathcal{A}_x(i,1)=\mathcal{A}_x(i,N_y)                \\
\mathcal{A}_y(i,1)=\mathcal{A}_y(i,N_y)                 \\

\end{array}
\end{equation}

\vspace{-0.5cm}

\section{3.Results}
\subsection{3.1.Domains And Spontaneous Vortex-Antivortex Pair}
In our calculation, the parameters for the lattice has been chosen as $N_x=N_y$ = 61 ,$d_x=d_y=\frac{16}{60} \xi_1 $. The relaxation step-size has been set as $\epsilon=0.01 $, the equation is iterated for one million steps.

Figure 1  records the variance of energy, after ten thousand steps, the change becomes very slight. Figure 2 records the max of $\frac{\partial{   \mathcal{F}}}{\partial{p} } $ which shows the speed of convergence, and found the max of $\frac{\partial{   \mathcal{F}}}{\partial{p} } $is less than $10^{-7}$ which meets the criteria for convergence and stablity. Here p stands for($ \widetilde{\psi_1}^*(i,j),\widetilde{\psi_2}^*(i,j), \mathcal{A}_x(i,j),\mathcal{A}_y(i,j)$) \\

\begin{figure}[H]
\centering 
\includegraphics[width=0.4\textwidth]{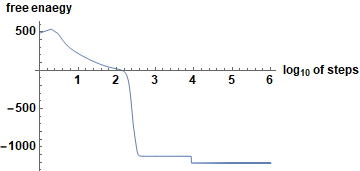}
\caption{(colour online)free energy vs. steps,parameters as $ a_2=1.2 $, $ b_2=a_2$, $c_2=1.1 $, $d_2=1.3$, $ \kappa=0.6$, $\Phi=0$. }
\end{figure}

\begin{figure}[H]
\centering 
\includegraphics[width=0.4\textwidth]{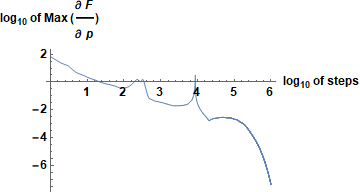}
\caption{(colour online)error  vs. steps,parameters as $ a_2=1.2 $, $ b_2=a_2$, $c_2=1.1 $, $d_2=1.3$, $ \kappa=0.6$, $\Phi=0$.  }
\end{figure}

From Figure 3, we can find that both the spontaneous magnetic B-field and wavefunction have the $C_4$ symmetry. It is quite nature, because the total free energy interaction between vortex-antivortex pairs are linear. Due to this symmetry, we can say that each vortex (antivortex)  exerts same force to its antivortex (vortex) neighbor, which can also be considered as each vortex trapped by their antivortex neighbor and vice versa, therefore the system will go back to its stable equilibrium if there are some perturbation.    \\

From Figure 3e/f and 4a/b, it is evident that our result can  be  analogous to the soliton behavior in two-band superconductor\cite{tanaka2001soliton}, where the sign of $\gamma$($d_2$ has the same sign as $\gamma $) specifies the strength of the interband interaction. When it is positive(Figure 3e/f), the relative phase of $ \psi_1 $ and $ \psi_2 $ is $\pi $. When it is negative(Figure 4a/b), the relative phase of $ \psi_1 $ and $ \psi_2 $ is $\pi $ is 0. This shows the influence of Josephson-coupling on the two-band model.

\vspace{-2cm}
\begin{figure}[H]
\centering  
\subfloat[(colour online)\ contourplot of the amplitude of$|\psi_1|^2$,the first order parameter]{%
\includegraphics[scale=0.27]{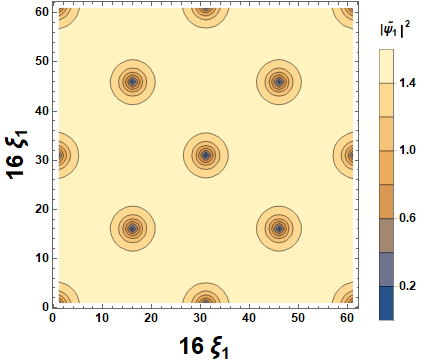}}\hfill
\subfloat[(colour online)\ contourplot of the amplitude of$|\psi_2|^2$,the second order parameter ]{%
\includegraphics[scale=0.27]{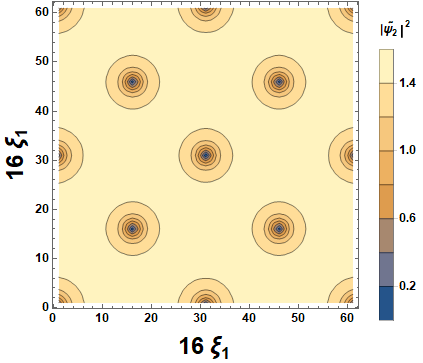}}\hfill

\subfloat[(colour online)\ 3D-plot of the amplitude of$|\psi_1|^2$,the first order parameter]{%
\includegraphics[scale=0.27]{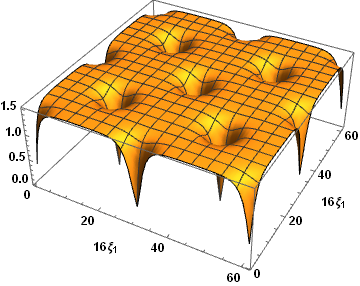}}\hfill
\subfloat[(colour online)\ 3D-plot of the amplitude of$|\psi_2|^2$,the second order parameter ]{%
\includegraphics[scale=0.27]{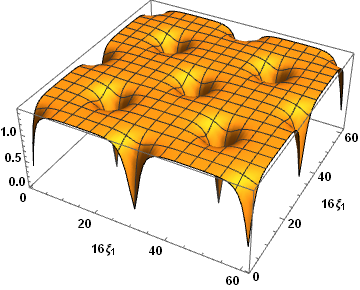}}\hfill

\subfloat[(colour online)\ vectorplot of the phase of $\psi_1$]{%
\includegraphics[scale=0.27]{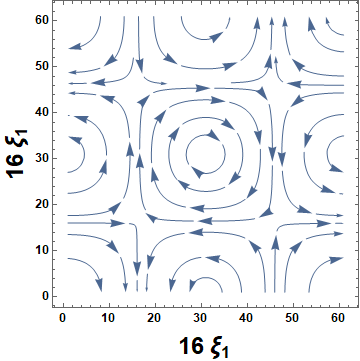}}\hfill
\subfloat[(colour online)\ vectorplot of the phase of $\psi_2$  ]{%
\includegraphics[scale=0.27]{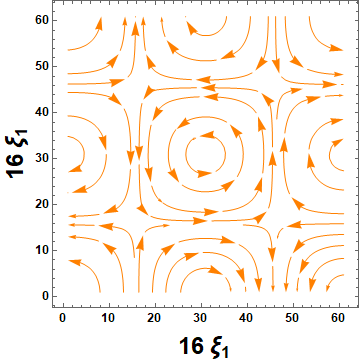}}\hfill

\subfloat[(colour online)\ contourplot of the spontaneous magnetic B-field ]{%
\includegraphics[scale=0.27]{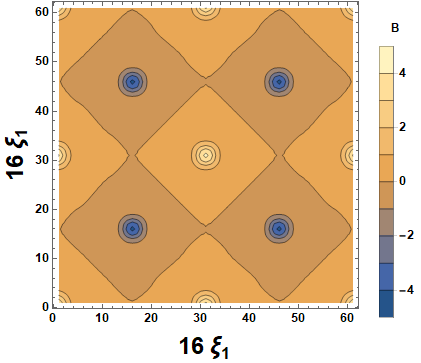}}\hfill
\subfloat[(colour online)\ contourplot of the  the amplitude of$|A|^2$,vector A-field ]{%
\includegraphics[scale=0.27]{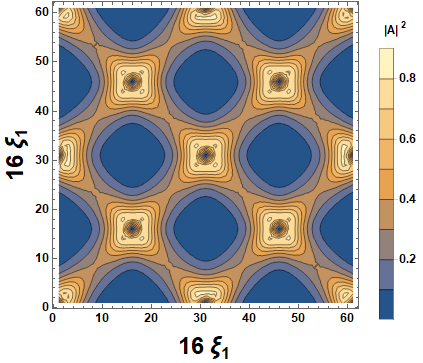}}\hfill

\subfloat[(colour online)\ 3D-plot of the spontaneous magnetic B-field ]{%
\includegraphics[scale=0.27]{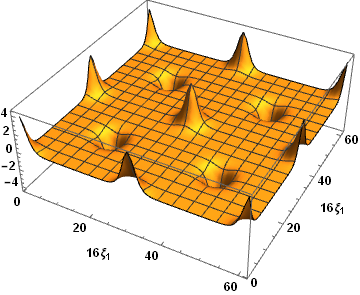}}\hfill
\subfloat[(colour online)\ vectorplot of the phase of $ A $ ]{%
\includegraphics[scale=0.27]{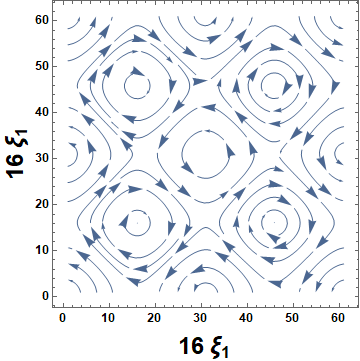}}\hfill

\caption{wavefunction,B-field and A-field,parameters as $ a_2=1.2 $, $ b_2=a_2$, $c_2=1.1 $, $d_2=1.3$, $ \kappa=0.6$, $\Phi=0$.}
\end{figure}

\begin{figure}[H]
\centering  
\subfloat[(colour online)\ contourplot of the amplitude of$|\psi_1|^2$, the first order parameter]{%
\includegraphics[scale=0.27]{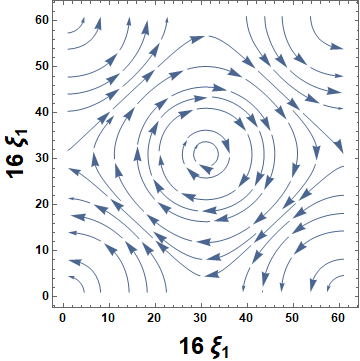}}\hfill
\subfloat[(colour online)\ contourplot of the amplitude of$|\psi_2|^2$, the second order parameter ]{%
\includegraphics[scale=0.27]{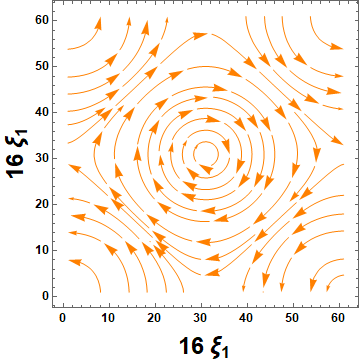}}\hfill

\caption{the phase of wavefunction,parameters as $ a_2=1.2 $, $ b_2=a_2$, $c_2=1.1 $, $d_2=-1$, $ \kappa=0.6$, $\Phi=0$.}
\end{figure}

The influence of Josephson-coupling on two-band model can also be displayed by the spontaneous magnetic B-field, there are four (as far we found) possible stable structure under different $d_2$ chosen.

\vspace{-0.5cm}

\begin{figure}[H]
\centering  

\subfloat[(colour online)\ contourplot of the spontaneous magnetic B-field under $d_2 $ = -1.]{%
\includegraphics[scale=0.28]{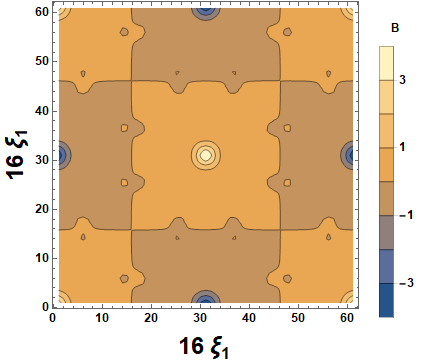}}\hfill
\subfloat[(colour online)\ contourplot of the spontaneous magnetic B-field under $d_2 $ =0.]{%
\includegraphics[scale=0.28]{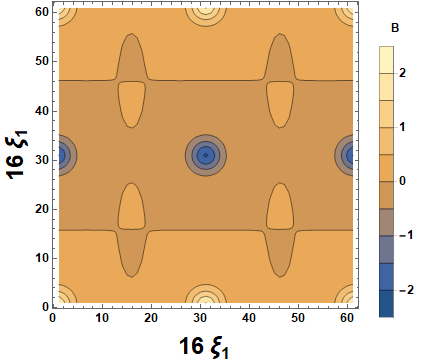}}\hfill

\subfloat[(colour online)\ contourplot of the spontaneous  magnetic B-field under $d_2 $ =1.]{%
\includegraphics[scale=0.28]{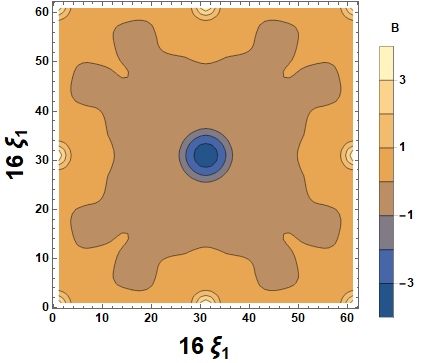}}\hfill
\subfloat[(colour online)\ contourplot of the spontaneous  magnetic B-field under $d_2 $ =1.3.]{%
\includegraphics[scale=0.28]{B.png}}\hfill

\caption{spontaneous magnetic B-field,parameters as $ a_2=1.2 $, $ b_2=a_2$, $c_2=1.1 $, $ \kappa=0.6$, $\Phi=0$  with different $d_2 $ chosen.}
\end{figure}

The aforementioned behavior is quite different from the behavior of one-band superconductor, where the vortex-antivortex pair can also appear spontaneously, but decay to blank pattern by the collision of pairs after several steps, the spontaneous pair keeps stable in the two-band situation.

\subsection{3.2.Stability of Vortex-antivortex Pair Under Applied B-field}

\vspace{-0.2cm}

To study the stability of vortex-antivortex pair in two-band model with Josephson-coupling, we apply magnetic B-field as perturbation. Under the influence of magnetic B-field,the spin should take the same direction as the applied field. From Figure 6, we found that the interband coupling is beneficial to the stability of the vortex-antivortex pair, since the stronger the intraband strength is, the longer transient state exists. And, Figure 7a/b/c/d shows the details of four stairs which witnesses four collisions between vortex and antivortex.

\vspace{-0.2cm}

\begin{figure}[H]
\centering 
\includegraphics[width=0.5\textwidth]{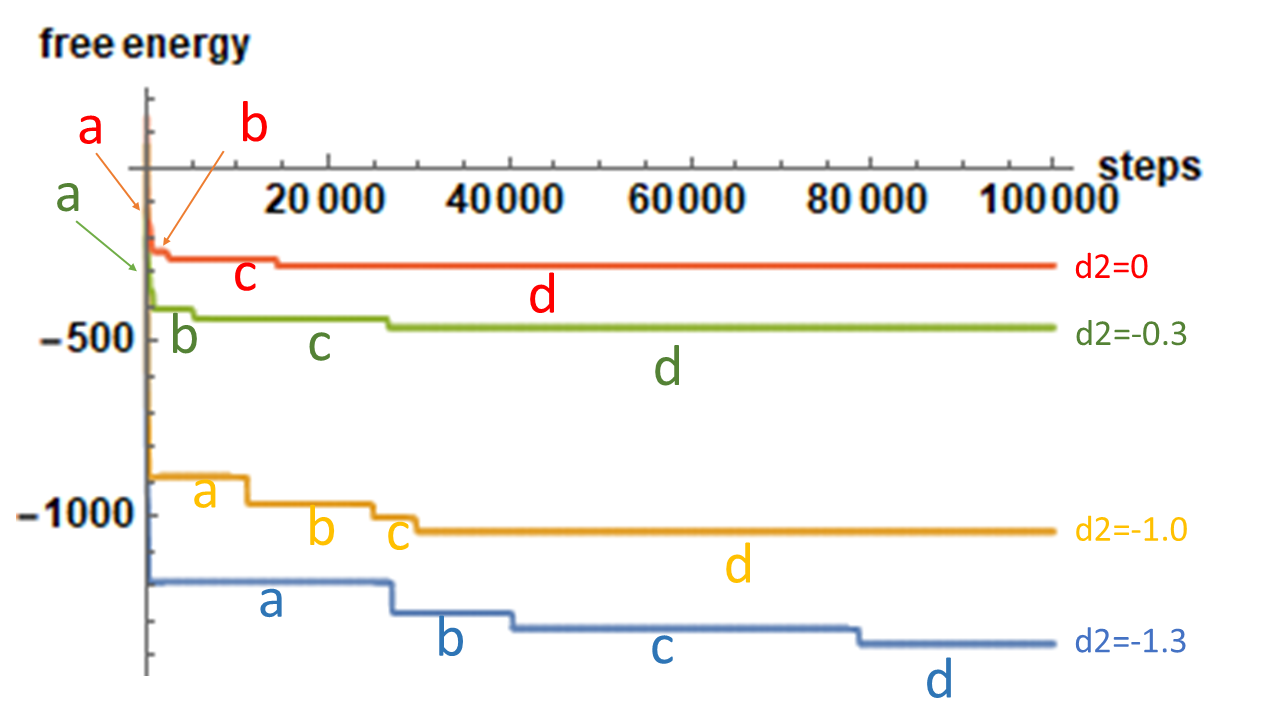}
\caption{(colour online)free energy vs. steps, parameters as $ a_2=1.2 $, $ b_2=a_2$, $c_2=1.1 $, $ \kappa=0.6$, $\Phi=2\pi$ with different $d_2$ chosen. Here, four stairs witnesses  four collisions. }
\end{figure}

\vspace{-1cm}

\begin{figure}[H]
\centering  
\subfloat[(colour online) steps=10000.]{%
\includegraphics[scale=0.29]{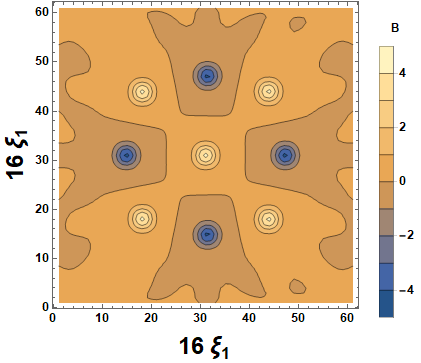}}\hfill
\subfloat[(colour online)steps=30000.]{%
\includegraphics[scale=0.29]{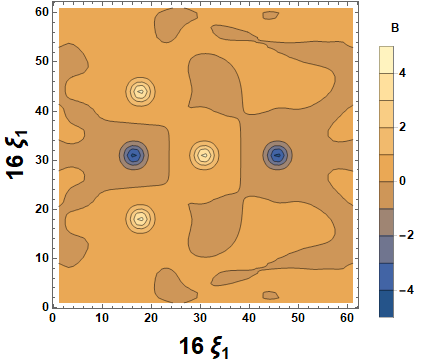}}\hfill

\subfloat[(colour online) steps=60000.]{%
\includegraphics[scale=0.29]{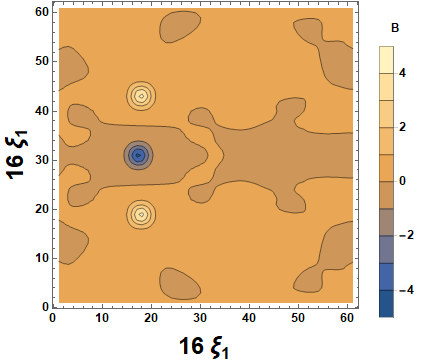}}\hfill
\subfloat[(colour online) steps=90000.]{%
\includegraphics[scale=0.29]{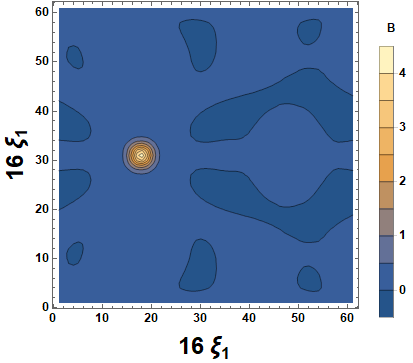}}\hfill

\caption{the frame of vortex-antivortex collision, parameters as $ a_2=1.2 $, $ b_2=a_2$, $c_2=1.1 $, $d_2=-1.3$, $ \kappa=0.6$, $\Phi=2\pi$.}
\end{figure}

\section{4.Conclusion}
We investigate the two-band GL model with Josephson-coupling on the basis of process of minimization the free energy with periodic boundary. A stable spontaneous  long-range vortex-antivortex induced by the Josephson-coupling, which changes the coherence behavior between two band, has been found by the  numerical search and theoretical analysis. In the meantime, this work shows the possibility of a new kind of topological excitations induced by the Josephson-coupling effect. Besides, there is an open question whether the similar spontaneous vortex-antivortex pair appears among other system with Josephson-coupling term or not. Furthermore, a generalization of this model by  comparison with bright dark soliton pure mathematics structure  is worth studying. Last but not least, this results shows possibility to trap unpaired majorana fermion if a two-band  $p_x+ip_y$ model (also with Josephson-coupling) applied.

\section{5.Acknowledgements}
We are grateful to Professor Shiping Zhou for his colourful discussions.

\section{\label{sec:level1}Reference}

\bibliography{bibfile}

\clearpage

\end{document}